\documentclass[pra,twocolumn,nofootinbib,eqsecnum,floatfix,superscriptaddress]{revtex4}
\usepackage{epsfig}
\usepackage{bm}
\usepackage{amsmath}
\input epsf
\numberwithin{equation}{section}
\renewcommand{\theequation}{\arabic{section}.\arabic{equation}}

\def\ed{ed.~by~}

\begin{document}

\title{Quasiclassical Coarse Graining \\ and Thermodynamic Entropy\footnote{Dedicated to Rafael Sorkin on his 60$^{th}$ birthday.}}

\author{Murray Gell-Mann}

\email{mgm@santafe.edu}

\affiliation{Santa Fe Institute, Santa Fe, NM 87501}

\author{James B.~Hartle}

\email{hartle@physics.ucsb.edu}

\affiliation{Santa Fe Institute, Santa Fe, NM 87501}
\affiliation{Department of Physics, University of California,
 Santa Barbara, CA 93106-9530}

\date{\today }

\begin{abstract}

Our everyday descriptions of the universe are highly coarse-grained, following only a 
tiny fraction of the variables necessary for a perfectly fine-grained description. 
Coarse graining in classical physics is made natural by our
limited powers of observation and computation. But in the modern quantum mechanics of
closed systems, some measure of coarse graining is inescapable because there are no 
non-trivial, probabilistic, fine-grained descriptions.  This essay explores the 
consequences of that fact.  Quantum theory allows for various coarse-grained descriptions some of which are mutually 
incompatible. For most purposes, however, we are interested in the small subset of ``quasiclassical descriptions''  defined by ranges of values of averages over small volumes of densities of conserved quantities such as energy and momentum and approximately conserved quantities such as baryon number. The near-conservation of these quasiclassical quantities  results in approximate decoherence, predictability, and local equilibrium, leading to closed sets of equations of motion. In any description, information is sacrificed through the coarse graining that yields decoherence and gives rise to probabilities for histories.
In  quasiclassical descriptions, further information is sacrificed in exhibiting the emergent regularities summarized by classical equations of motion.
An appropriate entropy measures the loss of information. For a ``quasiclassical realm''   this is connected with the usual thermodynamic entropy as obtained from  statistical mechanics. 
It was  low for the initial state of our universe and has been increasing since.

\end{abstract}



\maketitle

\section{Introduction}

Coarse graining is of the greatest importance in theoretical science, for example in connection with the meaning of regularity and randomness and of simplicity and complexity \cite{GL04}.  Of course it is also central to statistical mechanics, where a particular kind of coarse graining leads to the usual physico-chemical  or thermodynamic entropy. We discuss here the crucial role that coarse graining plays in quantum mechanics, making possible the decoherence of alternative histories and enabling probabilities of those decoherent histories to be defined.  The ``quasiclassical realms'' that exhibit how classical physics applies approximately in this quantum universe correspond to coarse grainings for which the histories have probabilities that exhibit a great deal of approximate determinism. In this article we connect these coarse grainings to the ones that lead, in statistical mechanics, to the usual entropy. 

Our everyday descriptions of the world in terms of nearby  physical objects like tables and clouds are examples of very {\it coarse-grained} quasiclassical realms.  At any  one time, these everyday descriptions 
follow only a tiny fraction of the variables  needed for a 
fine-grained description of the whole universe; they integrate over the rest.  Even those variables that
are followed are not tracked continuously in time, but rather sampled only at a sequence 
of times\footnote{{ This is not the most general kind of coarse graining. One can have instead a probability distribution characterized by certain parameters that correspond to the expected values of particular quantities, as in the case of absolute temperature and the energy in a Maxwell-Boltzmann distribution \cite{GL04}.}}. 

A classical gas of a large number of particles in a box provides other excellent examples of fine and coarse graining. The exact positions and momenta of all the particles as a
function of time give the perfectly fine-grained description.  Useful  coarse-grained
descriptions are provided by dividing the box into cells and specifying the volume
averages of the energy, momentum, and particle number in each.  Coarse graining can also be applied to ranges of values of the variables followed. This is an example in a classical context of what we will call a {\it quasiclassical coarse graining}. As we will describe in
more detail below, under suitable  initial  conditions and with suitable choices for the volumes,
this coarse graining leads to a deterministic, hydrodynamic description of the material
in the box.  We need not  amplify on the utility of that.

In classical physics, coarse graining arises from practical considerations. It might be forced on us by our puny ability to collect, store, recall, and manipulate data. 
Particular coarse grainings may be distinguished by their utility.  But there is always  
available, {\it in principle}, for every physical system,  an exact fine-grained description that could be used to answer any question, whatever the limitations on using it in 
practice.

In the modern formulation of the quantum mechanics of a closed system --- a formulation based on histories --- the situation is different. The probability that either of two exclusive histories will occur has to be equal to the sum of the individual probabilities. This will be impossible in quantum mechanics unless the interference term between the the two is made negligible by coarse graining. That absence of interference is called {\it decoherence}.   There {\it is no (non-trivial)  
fine-grained, probabilistic description of the histories of a closed quantum-mechanical  system}.  Coarse graining is
therefore inescapable in quantum mechanics if  the theory is to yield any predictions at
all.  

Some of the ramifications of these ideas are the following:

\begin{itemize}

\item Quantum mechanics  supplies probabilities for the members of various  decoherent sets of
coarse-grained alternative histories --- different `realms' for short. Different realms are {\it  compatible} if each one can be fine-grained to yield the same realm. (We have  a special case of this when one of the realms is a coarse graining of the other.) Quantum mechanics also exhibits mutually {\it incompatible} realms 
for which there is no finer-grained decoherent set of which they are both coarse grainings. 

\item Quantum mechanics by itself does not favor one realm over another. However, we are interested for most purposes in the family of quasiclassical realms  underlying everyday experience --- a very small subset of the set of  all realms.  Roughly speaking, a quasiclassical realm corresponds to coarse graining which follows ranges of the values of variables that enter into the classical equations of motion. 

\item Having non-trivial probabilities for histories requires the sacrifice of information
through the coarse graining necessary for decoherence.

\item Coarse graining beyond that necessary for decoherence is needed to achieve the
predictability represented by the approximate deterministic laws governing the sets of histories that constitute the quasiclassical realms.

\item An appropriately defined entropy that is a function of time is a useful measure of the information lost through a coarse graining defining a quasiclassical realm.  

\item  Such a coarse graining, necessary for having both decoherence and approximate classical predictability,  is connected with the coarse graining that defines the familiar entropy of thermodynamics. 

\item The entropy defined by a  coarse graining associated with a quasiclassical realm must be sufficiently  low at early times in our universe (that is, low for the coarse graining in question)  so that it tends to increase in time and exhibit the
second law of thermodynamics. But for this  the relaxation time for the increase of entropy must also be long compared with the age of the universe. From this point of view the universe is very young. 

\item The second law of thermodynamics, the family of quasiclassical realms, and even probabilities of any kind are features of our universe that depend, not just on its quantum state and dynamics, but also crucially on the inescapable coarse graining necessary to define those features, since there is no exact, completely fine-grained description.

\end{itemize}

Many of these remarks are contained in our earlier work \cite{GH90a,GH90b,Har91a,GH93a,GH93b,Gel94, GH94,Har94b,GH95}   or are  implicit in it or are contained in the work of others e.g. \cite{Omn94, Gri02} and the references therein. They are developed more extensively here in order to emphasize the central and inescapable role played by coarse graining in quantum theory. What is  new  in this paper includes the detailed analysis of the triviality of decoherent sets of  completely fine-grained histories,  the role of  narrative,  and, most importantly, the central result that the kind of  coarse graining defining the quasiclassical realms in quantum theory is closely related to the kind of coarse graining defining the usual entropy of chemistry and physics. 

The reader who is familiar with our previous work and notation can immediately jump to Section III. But for those who are not we offer a very brief summary in the next section.

\section{The Quantum Mechanics of a Closed System}

This section gives a bare-bones account of some essential elements of the modern synthesis
of ideas characterizing  the quantum mechanics of closed systems \cite{Gri02, Omn94, Gel94}.

To keep the discussion manageable, we consider a closed quantum system, most generally the
universe, in the approximation that gross quantum fluctuations in the geometry of spacetime
can be neglected. (For the generalizations that are needed for quantum spacetime see
{\it e.g.} \cite{Har95c,Har06}.)  The closed system can then be thought of as a large (say
$\gtrsim$ 20,000 Mpc), perhaps expanding box of particles and fields in a 
fixed background spacetime.
Everything is contained within the box, in particular galaxies, planets, observers and
observed , measured subsystems, and any apparatus that measures
them.  This is the most general physical context for prediction.

The fixed background spacetime means that the notions of time are fixed and that the
usual apparatus of Hilbert space, states, and operators can be employed in a quantum
description of the system.  The essential theoretical inputs to the process of prediction are
the Hamiltonian $H$ governing evolution (which we assume for simplicity to be time-reversible\footnote{For the issues arising from the time asymmetry
of the effective low-energy theory of the elementary particles in our local neighborhood of
the universe, see {\it e.g.} \cite{GH93b}.}) and the initial quantum condition (which we assume to be a pure state\footnote{From the perspective of the time-neutral formulation of quantum theory (e.g. \cite{GH93b}), we are assuming also a final condition of ignorance.}  $|\Psi\rangle$).  These
are taken to be fixed and given. 

The most general objective of quantum theory is the prediction of the probabilities of
individual members of sets of coarse-grained alternative histories of the closed
system. For instance, we might be interested in alternative histories of the center-of-mass
of the earth in its progress around the sun, or in histories of the correlation between the
registrations of  a measuring apparatus and a  property of the subsystem. Alternatives at one moment of
time can always be reduced to a set of yes/no questions.  For example, alternative positions 
of the earth's center-of-mass can be reduced to asking, ``Is it in this region -- yes or no?'',
``Is it in that region -- yes or no?'', etc. An exhaustive set of yes/no alternatives is
represented in the Heisenberg picture by an exhaustive set of orthogonal projection operators 
$\{P_\alpha(t)\}$,
$\alpha = 1, 2, 3 \cdots$.  These satisfy
\begin{equation}
\sum\nolimits_\alpha P_\alpha(t) = I, \  {\rm and}\ P_\alpha(t)\, P_\beta (t) =
\delta_{\alpha\beta} P_\alpha (t) \ ,
\label{twoone}
\end{equation}
showing that they represent an exhaustive set of exclusive alternatives.  In the Heisenberg
picture, the operators $P_\alpha(t)$ evolve with time according to 
\begin{equation}
P_\alpha(t) = e^{+iHt/\hbar} P_\alpha(0)\, e^{-iHt/\hbar}\, .
\label{twotwo}
\end{equation}
The state $|\Psi\rangle$ is unchanging in time.

Alternatives at one time can be described by expressing their projections in terms of fundamental
coordinates, say a set of quantum fields and their conjugate momenta.  In the Heisenberg
picture these coordinates evolve in time.  A given projection can be described in terms of
the coordinates at any time.  Thus, at any time it  represents  some alternative \cite{GH94}.

An important kind of set of exclusive histories is specified by sets of alternatives at a sequence of times $t_1<t_2<\cdots < t_n$. An individual history $\alpha$ in such a set is a particular sequence of alternatives  $\alpha \equiv (\alpha_1, \alpha_2, \cdots, \alpha_n)$. Such a set of histories has a branching structure in which a history up to any given time $t_m \le t_n$ branches into further alternatives at later times. `Branch' is thus an evocative synonym for such a  history\footnote{ Sets of histories defined by sets of alternatives at a sequence of discrete times is the simplest case sufficient for our purposes. For more general continuous time histories see e.g. \cite{Har91b,Ishsum}.}.

In such sets, we denote a projection at time $t_k$ by 
\begin{equation}
P^k_{\alpha_k} (t_k; \alpha_{k-1}, \cdots, \alpha_1)\, .
\label{fivetwo}
\end{equation}
We now explain this notation. 

Alternatives at distinct times can differ and are distinguished by the
superscript $k$  on the $P$'s. For instance, projections on ranges of position at one time  might be followed by projections on ranges of momentum at the next time, etc.

In realistic situations the alternatives will be  {\it branch-dependent}. In a
branch-dependent set of histories the {\it set} of alternatives at one time depends on the
particular branch. In \eqref{fivetwo} the string $\alpha_{k-1}, \cdots \alpha_1$ indicates this branch dependence. 
For example, in describing the evolution of the earth, starting with a protostellar cloud, a
relatively coarse-grained description of the interstellar gas might be appropriate in the
beginning, to be followed by finer and finer-grained descriptions at one location on the
branch where a star (the sun) condensed, where a planet (the earth) at 1AU won the battle of accretion in the circumstellar disk, etc. --- all events which happen only with some probability. Adaptive mesh refinement in numerical simulations of hydrodynamics provides a somewhat analogous situation.

An individual history $\alpha$  corresponding to a particular sequence of alternatives $\alpha \equiv (\alpha_1, \alpha_2, \cdots, \alpha_n)$  is represented by the corresponding chain of projections $C_\alpha$,  called a {\it class operator}. In the full glory of our notation this is:
\begin{equation}
C_\alpha\equiv P^n_{\alpha_n} (t_n; \alpha_{n-1}, \cdots , \alpha_1 ) \cdots  P^{2}_{\alpha_2} (t_2; \alpha_1) P^1_{\alpha_1} (t_1)\, .
\label{twothree}
\end{equation}
To keep the notation manageable we will sometimes not indicate the branch dependence explicitly where confusion is unlikely.

Irrespective of branch dependence, a set of histories like one specified by \eqref{twothree} is generally {\it coarse-grained}  because alternatives are specified
at some times and not at every time and because the alternatives at a given time are typically 
projections on subspaces with dimension greater than one and not projections onto a complete
set of states. Perfectly  {\it fine-grained} sets of histories consist of one-dimensional
projections at each and every time.

Operations of fine and coarse graining may be defined  on sets of histories.  A set of histories $\{\alpha\}$ may
be {\it fine -grained} by dividing up each class into
 an exhaustive set of exclusive subclasses $\{ \alpha' \}$.  Each subclass  consists of some histories in a coarser-grained class, and
every finer-grained subclass is in some  class.   {\it Coarse graining} is the operation of uniting subclasses of histories into bigger classes.  Suppose, for example, that the position
of the earth's center-of-mass is specified by dividing space into cubical regions of a
certain size. A coarser-grained description of position could consist of
larger regions made up of unions of the smaller ones. Consider a set of histories with class operators $\{C_\alpha \}$ and a coarse graining with class operators
 $\{ {\bar C}_{\bar\alpha} \}$  . The operators $\{{\bar C}_{\bar\alpha}\}$  are then related to the  operators $\{C_\alpha\}$ by summation, {\it viz.}
\begin{equation}
\bar C_{\bar\alpha} = \sum_{\alpha\in\bar\alpha} C_\alpha \ ,
\label{twofour}
\end{equation}
where  the sum is over the $C_\alpha$ for all finer-grained histories $\alpha$  contained within $\bar\alpha$.

For any individual history $\alpha$, there is a {\it branch state vector}  defined by
\begin{equation}
|\Psi_\alpha\rangle = C_\alpha |\Psi\rangle\, .
\label{twofive}
\end{equation}
When probabilities can be consistently assigned to the individual histories in a set,
they are given by
\begin{equation}
p(\alpha) = \parallel |\Psi_\alpha\rangle\parallel^2 = 
\parallel C_\alpha |\Psi\rangle\parallel^2\, .
\label{twosix}
\end{equation}
However, because of quantum interference,  probabilities cannot be consistently assigned to every set of alternative
histories that may be described.  The two-slit
experiment provides an elementary example: An electron emitted by a source can pass
through either of two slits on its way to detection at a farther screen.  It would be
inconsistent to assign probabilities to the two histories distinguished by which slit the
electron goes through if no ``measurement'' process determines this.  Because of interference, the probability for arrival  at a point
on the screen would not be the sum of the probabilities to arrive there by going
through each of the slits. In quantum theory, probabilities are squares of amplitudes
and the square of a sum is not generally the sum of the squares.

Negligible interference between the branches of a set
\begin{equation}
\langle\Psi_\alpha|\Psi_\beta\rangle\approx 0 \quad , \quad \alpha\not=\beta \, ,
\label{twoseven}
\end{equation}
is a sufficient condition for the probabilities \eqref{twosix} to be consistent with the
rules of probability theory. The orthogonality of the branches is approximate in
realistic situations. But we mean by \eqref{twoseven} equality to an accuracy that defines probabilities well beyond the
standard to which they can be checked or, indeed, the physical situation modeled
\cite{Har91a}. 

Specifically, as a consequence of \eqref{twoseven}, the probabilities \eqref{twosix} obey
the most general form of the probability sum rules
\begin{equation}
p(\bar\alpha) \approx \sum_{\alpha\in\bar\alpha} p(\alpha)
\label{twoeight}
\end{equation}
for any coarse graining $\{{\bar\alpha}\}$ of the $\{\alpha\}$.  Sets of
histories obeying \eqref{twoseven} are said to (medium) decohere. As L.~Di\'osi has shown \cite{Dio04}, medium
decoherence is the weakest of known conditions that are consistent with elementary notions of the
independence of isolated systems\footnote{For a
discussion of the linear positive, weak, medium, and strong decoherence conditions,
see \cite{GH90b, GH95, Har04}.}. 
Medium-decoherent sets are thus the ones for which quantum mechanics consistently makes predictions of probabilities through 
\eqref{twosix}.  Weaker conditions, such as those defining merely ``consistent'' histories, are not appropriate. 

The decoherent sets exhibited by our universe are determined through \eqref{twoseven} and by the Hamiltonian $H$ and the quantum state $|\Psi\rangle$.
We use the term {\it realm} as a synonym for a decoherent set of coarse-grained
alternative histories.

A coarse graining of a decoherent set is again decoherent. A fine-graining of a
decoherent set risks losing decoherence. 

It is more general to allow a density matrix $\rho$ as the initial quantum state.
Decoherence is then defined in terms of a {\it decoherence functional}. 
\begin{equation}
D(\alpha, \beta) = Tr\, (C_\alpha \rho C^\dagger_\beta)\, .
\label{twonine}
\end{equation}
The rules for both decoherence and the existence  of  probabilities can then be expressed in a single
formula
\begin{equation}
D(\alpha, \beta) \approx \delta_{\alpha\beta} p (\alpha)\, .
\label{twoten}
\end{equation}
When the density matrix is pure, $\rho = |\Psi\rangle\langle\Psi|$, the condition
\eqref{twoten} reduces to the decoherence condition \eqref{twoseven} and the
probabilities to \eqref{twosix}. 

An important mechanism of decoherence is the dissipation of phase coherence between
branches into variables not followed by the coarse graining.  Consider by way of
example, a dust grain in a superposition of two positions deep in interstellar space
\cite{JZ85}.  In our universe, about $10^{11}$ cosmic background photons scatter from
the dust grain each second.  The two positions of the grain become correlated with different, nearly
orthogonal states of the photons. Coarse grainings that follow only the position of the
dust grain at a few times therefore correspond to branch state vectors that are nearly
orthogonal and satisfy \eqref{twoeight}.  

Measurements and observers play no fundamental role in this general formulation
of usual quantum
theory.  The probabilities of measured outcomes can be computed and are given to
an excellent approximation by the usual story. But, in a set of histories where they
decohere, probabilities can be assigned to the position of the moon when it is not
receiving the attention of observers and to the values of density fluctuations in the early 
universe when there were neither measurements taking place nor observers to carry them out.

The probabilities of the histories in the various realms  and the conditional probabilities constructed from them
constitute  the predictions of the quantum mechanics of a closed
system given the Hamiltonian $H$ and initial state $|\Psi\rangle$.

\section{Inescapable Coarse Graining}

In this section and Appendix A, we show that there are no exactly
decoherent, completely fine-grained sets of alternative histories describing a closed quantum 
system except trivial ones which essentially reduce to a description at only one time\footnote{ The observation that coarse graining is necessary for non-trivial decoherence  has been made since the start of decoherent histories quantum theory. See, e.g. \cite{GH90a,Dio95,DK96,Cra97} for some different takes on this.}  (In Appendix B,  we
also show that there is no certainty except that arising from the unitary evolution controlled by the Hamiltonian.)

A set of alternatives at one time is completely  fine-grained if its projections are onto an
orthonormal basis for Hilbert space.  Specifically,
\begin{equation}
P_i (t) = |i\rangle\langle i|
\label{threeone}
\end{equation}
where $\{|i\rangle\}\, , i=1, 2, \cdots$ are a set of orthonormal basis vectors. Any basis defines at any time
some set of alternatives \cite{GH94}.

Probabilities are predicted for any fine-grained set of alternatives at one time.  The
branch state vectors are
\begin{equation}
|\Psi_i\rangle = P_i(t)|\Psi\rangle = |i\rangle\langle i|\Psi\rangle\, .
\label{threetwo}
\end{equation} 
These are mutually orthogonal and therefore decohere [cf. \eqref{twoseven}]. The consistent
probabilities are 
\begin{equation}
p(i) = |\langle i |\Psi\rangle |^2\, .
\label{threethree}
\end{equation}

A completely fine-grained description does not consist merely of fine-grained alternatives
at one time but rather of fine-grained alternatives at each and every available time.  
Specifically, a
completely fine-grained set of histories is a set of alternative histories specified by
sets of fine-grained alternatives like \eqref{threeone} at every available time.

To avoid the mathematical issues that arise in defining continuous infinite products of
projections [{\it cf.}~\eqref{twothree}], we will assume that the available times are
restricted to a large discrete series $t_1, \cdots, t_n$ in a fixed interval $[0, T]$.
Physically there can be no objection to this if the intervals between times is taken
sufficiently small (if necessary of order the Planck time).  More importantly, the absence of
decoherent sets with a finite number of times prohibits the existence of any finer-grained decoherent sets,
however they are defined.

A completely fine-grained set of histories is then specified by a series of bases $\{|1,
i_1\rangle\}, \{|2, i_2\rangle\}, \cdots, \{|n, i_n\rangle\}$ defining one dimensional
projections of the form \eqref{threeone}. In $|k, i_k\rangle$ the first argument labels
the basis, the second the particular vector in that basis.  
To get at the essentials of the argument, we
will assume in this section the generic situations where there are no branch state
vectors that vanish. More particularly, we assume that none of the matrix elements
$\langle k+1, i_{k+1} |k, i_k\rangle$ or $\langle 1, i_1|\Psi\rangle$ vanishes.  Put
differently, we assume that none of the questions at each time is trivially related to a
question at the next time or to the initial state.  The general case where these
assumptions are relaxed is dealt with in Appendix A along with the related notion of trivial
decoherence that  arises.

The branch state vectors for the histories $(i_1, \cdots, i_n)$ in such a completely
fine-grained set have the form
\begin{equation}
|n, i_n\rangle\langle n, i_n | n-1, i_{n-1}\rangle \cdots \langle 2,
i_2|1,i_1\rangle\langle1, i_1|\Psi\rangle\, .
\label{threefour}
\end{equation}
 The vectors $|i_n,n\rangle$  are orthogonal for different $i_n$.  The condition for decoherence \eqref{twoseven}  then requires
\begin{eqnarray}
\langle\Psi |1, i^\prime_1\rangle\langle1, i^\prime_1  |2, i^\prime_2\rangle \cdots  \langle
n, i_n|n, i_n\rangle \cdots  \nonumber \\ \times 
\langle 2, i_2|1, i_1\rangle\langle 1,
i_1|\Psi\rangle=0
\label{threefive}
\end{eqnarray}
whenever any $i^\prime_k\not= i_k$ for $k=1, \cdots, n-1$. But, by assumption, none of these matrix elements
vanishes so there are no exactly decoherent, completely fine-grained sets of this kind.  

To get a different perspective, suppose that $\langle 1, i_1|\Psi\rangle$ is allowed
to vanish for some $i_1$ but the restriction that $\langle k, i_k|k-1,
i_{k-1}\rangle\not= 0$ is retained.  Then \eqref{threefive} could be satisfied
provided there is {\it only one} $i_1$ for which $\langle 1, i_1|\Psi\rangle\not=0$. But
since $\{|1, i_1\rangle\}$ is a basis, this means it must consist of $|\Psi\rangle$ and a
complete set of orthogonal vectors.  But that means that the first set of alternatives is
trivial.  It merely asks, ``Is the state $|\Psi\rangle$ or not?''  As explained in
Appendix A, there are no exactly decoherent, completely fine-grained sets of alternative
histories of a closed quantum system that are not trivial in this or similar senses.
Coarse graining is therefore inescapable.

\section{Comments on  Realms}

A perfectly  fine-grained set of histories can be coarse-grained in many different ways to
yield a decoherent set whose probabilities can be employed in the process of
prediction. Furthermore, there are many different completely fine-grained sets to start
from, corresponding to the possible choices of the bases $\{|k, i_k\rangle\}$ at each
time arising from different complete sets of commuting observables.  Once these fine-grained sets are 
coarse-grained enough to achieve decoherence, further coarse graining
preserves decoherence.  Some decoherent sets can be organized into compatible families all members of which are  coarse grainings of a common finer-grained decoherent set.  But there remain distinct families of decoherent sets
which have no finer-grained decoherent sets of which they are all coarse grainings. 
As mentioned in the Introduction, these are called
{\it incompatible} realms. 

 We may not draw inferences by combining
probabilities from incompatible realms.\footnote{It has been shown, especially by R.~Griffiths
\cite{Gri02}, that essentially all inconsistencies alleged against consistent
histories quantum mechanics (or, for that matter, decoherent histories quantum mechanics)  arise from violating this logical prohibition.} 
That would implicitly assume, contrary to fact, that
probabilities of a finer-grained description are available. Incompatible
decoherent sets provide different, incompatible descriptions of the universe.  Quantum theory does not automatically prefer one of these realms over another without further criteria such as quasiclassicality.  

Note that in{\it compatibility} is not in{\it consistency} in the sense of predicting different probabilities for the same histories in different realms. The probability for a history $\alpha$ is given by \eqref{twosix}  in any realm of which it is a member. 

While quantum theory permits a great many incompatible realms for the description of a closed system, we as
observers utilize mainly sets that are coarse grainings of one family of such realms ---
the quasiclassical realms  underlying everyday experience.\footnote{Some authors, notably Dowker and Kent \cite{DK96}, have suggested that quantum mechanics is incomplete without a
fundamental set selection principle that would essentially single out the quasiclassical
realms from all others.  The following discussion of the properties of quasiclassical
realms could be viewed as steps in that direction. We prefer to keep the fundamental 
formulation clean and precise  by basing it solely on the notion of decoherence.}  We now turn to a characterization of those.

The histories of a quasiclassical realm  constitute {\it narratives}. That is,
they describe how certain features of the universe change over time. A narrative
realm is a particular kind of set of histories in which the projections at successive
times are related by a suitable rule. That way the histories are stories about what 
happens over time and not simply unrelated or redundant scraps of information from a sequence of times.

The simplest way of defining a narrative is to take the same quantities at each time;  histories then describe how those quantities change over time. The corresponding rule connecting the projections at different times can be simply stated in the Schr\"odinger picture, where operators corresponding to a given quantity do not change in time.  The set of Schr\"odinger picture projections $\{{\hat P}_\alpha\}$ is the same at each time. Equivalently, in the Heisenberg picture the projections at each time are given by
\begin{equation}
P^k_\alpha (t_k) = e^{iHt_k/\hbar}\, \hat P_\alpha e^{-iHt_k/\hbar} \ .
\label{fiveone}
\end{equation}
The model coarse graining using the same  quasiclassical variables at each time step that will be presented in the next section obeys \eqref{fiveone}.  But in a more realistic situation the histories will be more complicated, especially because they exhibit branch dependence.

There are trivial  kinds of non-narrative realms with a high level of
predictability but little or no utility.  Examples are realms that mindlessly repeat the same
set of Heisenberg picture projection operators at each and every time, and the trivial
fine-grained realms described in Section III.A and Appendix A.  Suppose, for instance, the same Heisenberg picture projection operators are repeated at each time so that $P^k_{\alpha_k}(t_k) = Q_{\alpha_k}$  for some set of orthogonal projections $\{Q_\alpha \}$.  This leads to the following form for the $C_\alpha$: 
\begin{equation}
C_\alpha = Q_{\alpha_n} \cdots  Q_{\alpha_1} \, . 
\label{fiveonea}
\end{equation}
The only non-vanishing $C$'s are just the projections $Q_\alpha$.  Effectively the histories are specified by alternatives at one time. Nothing happens!  Restricting to narrative realms eliminates this kind of triviality. 

It remains to specify carefully what is a suitable rule for relating the projection operators in each history at successive times, thus providing a general definition of a narrative realm. We are searching for the best way to do that, using ideas about simpliciy, complexity, and logical depth presented in \cite{GL04}.

\section{The Quasiclassical Realm(s)}

As discussed in Section II, coarse graining is necessary for probability. The families of decoherent sets of
coarse-grained histories give rise to descriptions of the universe that are often incompatible with one another. As information gathering and utilizing systems
(IGUSes), we use, both individually and collectively, only a very limited subset of these
descriptions belonging to a compatible  family of realms with histories and probabilities that manifest certain regularities of the universe  associated with classical dynamical laws. These regularities include ones that are exploitable in our various pursuits, such as getting food,  reproducing, avoiding becoming food,  and achieving  recognition.  
Such sets of histories are defined by alternatives that include ones which our perception is adapted to distinguish.  
As we will see, coarse graining very far  beyond that necessary for mere decoherence is necessary to 
define the sets that exhibit the most useful of these regularities.

Specific systems like the planet Mars, Western culture, and asparagus exhibit various kinds of
particular exploitable regularities.   But the most widely applicable class of regularities
consists of  the correlations in time governed by the deterministic laws of motion of classical
physics.  In our quantum universe, classical laws are approximately applicable over a wide range of times,
places, scales, and epochs.  They seem to hold approximately over the whole of the visible universe from a time shortly after the beginning to the present. Indeed, we expect them to hold into the far  future.  We refer to the family of decoherent sets of coarse-grained histories that
describe these regularities as the family of {\it quasiclassical realms}.

The characteristic properties of a quasiclassical realm all follow from the approximate conservation of the variables that define them, which we call {\it quasiclassical variables.} 
As we will illustrate with a specific model below, these include averages of densities of quantities such as energy and momentum that are exactly conserved\footnote{Their exact conservation reflects the local symmetries of the approximately fixed spacetime geometry that emerges from the quantum-gravitational fog near the beginning.}. But they also include densities of quantities such as baryon number that may be conserved only  to an approximation that varies with the epoch. 

Approximate conservation leads to predictability in the face of the noise that accompanies typical mechanisms of decoherence. Approximate conservation allows the local equilibrium that leads to closed sets of classical equations of motion summarizing that predictability. This local equilibrium is the basis for the definition of an entropy implementing the second law of thermodynamics.


More specifically, in this paper,  by a {\it quasiclassical realm} we mean an exhaustive set of mutually exclusive
coarse-grained alternative histories that obey  medium decoherence with the following additional properties: The histories 
consist largely of related but branch-dependent projections onto ranges  of quasiclassical variables at a succession of times.  Each history with a non-negligible probability constitutes a narrative, with
individual histories  exhibiting patterns of correlation implied by closed sets of effective
equations of motion interrupted by frequent small fluctuations and occasional major branchings  (as in measurement situations).  By a {\it family of quasiclassical realms} we mean a set of compatible ones that  are all  coarse grainings of a common one.   Useful families span an enormous range of coarse graining. Quasiclassical realms describing everyday experience, for instance, may be so highly coarse-grained as to refer merely to the features of a local environment deemed worthy of an IGUS's attention. At the other end of the range are the quasiclassical realms that are as fine-grained as possible given  decoherence and quasiclassicality.  Those realms extend over the wide range of time, place, scale and epoch mentioned above. These coarse grainings defining the maximally refined quasiclassical realms are not a matter of our choice. Rather, those maximal realms are a feature of our universe that we exploit --- a feature that is emergent from   the initial quantum state, the Hamiltonian, and the extremely long sequence of outcomes of chance events\footnote{In previous work we have taken the term `quasiclassical realm' to be defined in some respects more generally and  in other respects more restrictively than we have here \cite{GH90a}.  To investigate information-theoretic measures for quasiclassicality, we left open the possibility that there might be realms exhibiting deterministic correlations in time defined by variables different from the quasiclassical variables. The `quasiclassical realms' of this paper were called `usual quasiclassical realms'. We also required that every quasiclassical realm be maximal in the sense discussed above so that it was an emergent feature of the universe and not our choice. For the limited objectives of this paper it seems best to employ the simpler terminology that we have used here rather than seek exact consistency. We leave open whether there are  incompatible maximal realms that also exhibit high levels of predictability and utility. }.

\subsection{Quasiclassical Variables}

To build a simple model with a coarse graining that defines a quasiclassical realm, we begin with a large box containing local quantum fields.  The interactions of these fields are assumed to be local and, for simplicity,  to result in short-range forces on bulk matter.   Spacetime is assumed to be nearly flat inside the box so that large fluctuations in the geometry of spacetime are  neglected.  This is a
reasonably general model for much of physics in the late universe.

Space inside the box is divided into equal volumes of size $V$ labeled by a discrete
index $\vec y$. (The branch dependence of volumes described in Section II is thus ignored.) The linear dimensions of the volumes are assumed to be large compared to the ranges of the forces. Quasiclassical variables are constructed as averages over these volumes of
conserved or approximately conserved extensive quantities.  These will include energy and linear momentum
since spacetime is treated as  flat.  There might be many conserved or approximately conserved species of particles. To keep the
notation manageable we consider just one.

To be explicit, let $T^{\alpha\beta} (\vec x, t)$ be the  stress-energy-momentum 
operator for the quantum fields in the Heisenberg picture.  The energy density $\epsilon 
(\vec x, t)$ and momentum density $\pi^i (x, t)$ are $T^{tt}(\vec x, t)$ and $T^{ti} 
(\vec x,t)$ respectively.  Let $\nu (\vec x, t)$ denote the number density of the conserved or nearly conserved species. Then we define
\begin{subequations}
\label{sixone}
\begin{eqnarray}
{{\epsilon}_V}(\vec y, t) & \equiv & \frac{1}{V} \int_{\vec y} d^3x\, \epsilon (\vec x, t)
\label{sixone a}\\
\vec\pi_V (\vec y, t) & \equiv & \frac{1}{V} \int_{\vec y} d^3x
\vec\pi(\vec x, t)
\label{sixone b}\\
\nu_V (\vec y, t) & \equiv & \frac{1}{V} \int_{\vec y} d^3x\, \nu (\vec x, t)
\label{sixone c}
\end{eqnarray}
\end{subequations}
where in each case the integral is over the volume labeled by $\vec y$. These are the
quasiclassical variables for our model.  We note that the densities in \eqref{sixone}
are the variables for a classical hydrodynamic description of this system --- for example, the variables of the Navier-Stokes equation. In more general situations, these
might be augmented by further densities of conserved or nearly conserved species, field averages, and other variables, but we will restrict our  attention to just these. 

A set of alternative coarse-grained histories can be constructed by giving ranges of
these quasiclassical variables at a sequence of times. We will call these a set of quasiclassical
histories. To be a
quasiclassical realm, such a set must decohere and the probabilities must be high for
the correlations in time specified by a closed set of deterministic equations of
motion. In Section C we will review the construction of these sets of histories, their 
decoherence, and their probabilities, referring to  the work of Halliwell \cite{Hal98, Hal99,Hal03}. 
But here we anticipate the qualitative reasons for these results.

Typical realistic mechanisms of decoherence involve the dissipation of  phases (between the branch state vectors) into 
variables that are not  followed by the coarse graining. That is the case, for instance, with the many models in which the position of
one particle is followed while it is coupled to a bath of others whose positions are
ignored for all time\cite{JZ85, FV63, CL83, GH93a}. Quasiclassical coarse grainings  do not posit one set of variables that are ignored (integrated over)  for all time, constituting a fixed `environment'. Rather, at each time,  the projection operators of the coarse graining define the ignored interior configurations of the volumes whose overall energy, momentum, and number are
followed.\footnote{ As shown in \cite{BH99}, for any coarse graining of the form \eqref{twofour}  it is possible to factor an infinite-dimensional Hilbert space into a part that is followed by the coarse graining and a part that is ignored (an environment.) However, that factorization may change from one time to the next, making it difficult to use.
The linear oscillator chain studied in \cite{BH99} provides another example of decoherence brought about  by a coupling of followed variables to internal ones.} 

The coupling between followed and ignored variables that is necessary for decoherence
is inevitably a source of noise for the followed quantities, causing deviations from
predictability. The approximate conservation of quasiclassical variables allows them to
resist the noise that typical mechanisms of decoherence produce and to remain approximately predictable  because typically the inertia of each relevant degree of freedom is large \cite{GH93a,Har94b}.

Indeed, consider the limit where there is only one volume occupying the whole box.
Then the quasiclassical variables are the total energy, total momentum, and total
number of particles in the box. These are exactly conserved and mutually commuting.  Histories of these quantities are therefore
precisely correlated with the final values and trivially decohere in the sense discussed in Section III and Appendix A.  Exact decoherence and 
persistence in the limit of one volume suggest efficient approximate 
decoherence for smaller volumes.

Decoherence, however, is not the only requirement for a quasiclassical realm.  A
quasiclassical realm must also exhibit the correlations in time implied by a closed set
of classical equations of motion. It is to this property that we now turn.

\subsection{Classical Equations for Expected Values}

Isolated systems generally evolve toward equilibrium.  That is a consequence of statistics. But conserved or 
approximately conserved quantities such as energy, momentum, and number approach equilibrium more slowly than others. That means that a situation of {\it local equilibrium} will generally
be reached before complete equilibrium is established, if it ever is. This local equilibrium is characterized
by the values of conserved quantities constained in small volumes.  Even for systems of modest size, time
scales for small volumes to relax to local equilibrium can be very, very much shorter than the time scale for reaching
complete equilibrium\footnote{  In realistic situations there will generally be a hierarchy of time scales in which different kinds of equilibrium are reached on different distance scales. Star clusters provide a simple example. The local equilibrium of the kind described here is reached much more quickly  for the matter inside the stars than the metastable equilibrium  governed by weak, long range gravitation that may eventually characterize the cluster as a whole. By restricting our model to short-range forces we have avoided such realistic complications.}  Once local equilibrium is established, the subsequent evolution of
the approximately conserved quantities can be described by closed sets of effective
classical equations of motion such as the Navier-Stokes equation.  The local equilibrium
determines the values of the  phenomenological quantities such as pressure and viscosity that enter into these equations and the relations among them.  This section reviews the standard derivation of these equations of motion for the
expected values of the approximately conserved quantities  for our model universe in a box, as can be found for example in \cite{Fos75, Zub74}.
The next section considers the histories of these quantities, their decoherence, and their
probabilities.

Central to the description of local equilibrium is the idea of an {\it effective density
matrix}. For large systems the computation of the decoherence functional directly from the
quantum state $\rho \equiv|\Psi\rangle\langle\Psi|$ may be practically impossible. However, it may
happen for {\it certain classes of coarse grainings} that the decoherence functional for a coarse  graining  is given to a
good approximation by an effective density matrix $\tilde \rho$  that requires less information to specify than $\rho$ does (cf Section VI). That is,
\begin{equation}
D(\alpha, \beta) \equiv Tr (C_\alpha \rho C^\dagger_\beta) \approx 
Tr (C_\alpha \tilde\rho C^\dagger_\beta)
\label{sixtwo}
\end{equation}
for all histories $\alpha$ and $\beta$ in  the exhaustive set of alternative coarse-grained
histories\footnote[11]{ The $\approx$ in  \eqref{sixtwo} means equality to the extent that there is local equilibrium. This is a different approximation from that in \eqref{twoten}, where it is the off-diagonal elements of the decoherence functional that are negligible.}.

A familiar example of an effective density matrix is the one describing a system in a box in
thermal equilibrium
\begin{equation}
\tilde\rho_{\rm eq} = Z^{-1} \exp[-\beta (H-\vec U\cdot \vec P -\mu N)]\, .
\label{sixthree}
\end{equation}
Here, $H$, $\vec P$, and $N$ are the operators for total energy, total momentum, and total
conserved number inside the box --- all extensive quantities.  The $c$-number intensive
quantities $\beta$, $\vec U$, and $\mu$ are respectively the inverse temperature (in units
where Boltzmann's constant is 1), the velocity of the box, and the chemical potential.  A
normalizing factor $Z$ ensures $Tr(\tilde\rho_{\rm eq})=1$.  In the next section,  we will give a standard derivation of this effective density matrix as one that maximizes the missing information subject to expected value constraints.  

Local equilibrium is achieved when the decoherence functional for sets of histories of
quasiclassical variables $(\epsilon, \vec\pi, n)$ is given approximately by the {\it local} version of
the equilibrium density matrix \eqref{sixthree}  
\begin{align}
\tilde\rho_{\rm leq}=&Z^{-1} \exp\Bigl[-\sum\limits_{\vec y} \, \beta(\vec y, t)\, \bigl(\epsilon_V(\vec y,
t) \nonumber \\& - \vec u(\vec y, t)\cdot \vec \pi_V\, (\vec y, t) - \mu\, (\vec y, t)\, \nu_V\, (\vec y,
t)\bigr)\Bigr] \ . 
\label{sixfive}
\end{align}
This local equilibrium density matrix is constructed to reproduce the expected values of the quasiclassical variables such as the energy density averaged over a volume at a moment of time $\epsilon_V (\vec y, t)$, viz. 
\begin{equation}
\langle\epsilon_V (\vec y, t)\rangle \equiv Tr (\epsilon_V (\vec y, t) \rho) =Tr (\epsilon_V (\vec y, t) \tilde\rho_{\rm leq})\, 
\label{sixfour}
\end{equation}
and similarly for $\vec\pi_V (\vec y, t)$ and $\nu_V (\vec y, t)$.
The expected values of quasiclassical quantitites are thus functions of the 
intensive $c$-number quantities $\beta(\vec y, t)$, $\vec u(\vec y, t)$, and $\mu(\vec y,
t)$. These  are the local inverse temperature, velocity, and chemical potential respectively. They now vary with time as the system evolves toward complete equilibrium.  In the next section,  we will give a standard derivation of the local equilibrium density matrix as one that maximizes the missing information subject to expected value constraints of local quantities as in \eqref{sixfour}.


When there is not too much danger of confusion we will often replace sums over $\epsilon_V(\vec y, t)$ with integrals over $\epsilon(\vec x, t)$ and similarly with the other quasiclassical quantities $\vec\pi$ and $\nu$. In particular, the differential equations of motion that we discuss below are a familiar kind of  approximation to a set of difference equations.

A closed set of deterministic equations of motion for the expected values of $\epsilon (\vec x,
t)$, $\vec\pi (x, t)$, and $\nu(\vec x, t)$ follows from assuming that $\tilde\rho_{\rm leq}$
is an effective density matrix for computing them.  To see this, begin with the Heisenberg equations for the conservation of the  stress-energy-momentum operator $T^{\alpha\beta}(\vec x, t)$ and the number
current $j^\alpha(\vec x, t)$.
\begin{equation}
\frac{\partial T^{\alpha\beta}}{\partial x^\beta}=0\quad , \quad \frac{\partial
j^\alpha}{\partial x^\alpha}=0\, .
\label{sixsix}
\end{equation}
 Noting
that $\epsilon(\vec x, t)=T^{tt} (\vec x, t)$ and $\pi^i (\vec x, t)=T^{ti} (\vec x, t)$,
eqs \eqref{sixsix} can be written out in a 3$+$1 form and their expected values taken.  The result is the set of  five equations
\begin{subequations}
\label{sixseven}
\begin{eqnarray}
\frac{\partial\langle\pi^i\rangle}{\partial t} & = & -\frac{\partial\langle
T^{ij}\rangle}{\partial x^j}\, ,\label{sixseven a}\\
\frac{\partial\langle\epsilon\rangle}{\partial t} & = & -\vec\nabla\cdot\langle\vec\pi\rangle
\, ,\label{sixseven b}\\
\frac{\partial\langle\nu\rangle}{\partial t} & = & -\vec\nabla\cdot\langle\vec\jmath\rangle\, .
\label{sixseven c}
\end{eqnarray}
\end{subequations}
The expected values are all functions of $\vec x$ and $t$..

To see how these conservation laws lead to a closed system of equations of motion, consider the
right hand sides, for instance that of \eqref{sixseven a}. From the form of 
$\tilde\rho_{\rm leq}$
in \eqref{sixfive} we find that the stress tensor $\langle T^{ij} (\vec x, t)\rangle$ is a
{\it function} of $\vec x$ and $t$, and a {\it functional} of the intensive multiplier functions 
$\beta(\vec \xi, \tau)$, $\vec u(\vec\xi, \tau)$, and $\mu(\vec \xi, \tau)$. We write this relation as
\begin{equation}
\langle T^{ij} (\vec x, t)\rangle = \check T^{ij} [\beta(\vec\xi , \tau)\, , 
u(\vec\xi , \tau)\, ,
\mu(\vec\xi, \tau)\, ;\, \vec x, t)\, .
\label{sixeight}
\end{equation}
In the same way, the expected values of $\epsilon(\vec x, t)$, $\vec\pi(x, t)$, and
$\nu(\vec x, t)$ become functionals of $\beta(\vec\xi, \tau)$, $\vec u(\vec\xi, \tau)$, and $\mu(\vec\xi, \tau)$, {\it e.g.}
\begin{equation}
\langle\epsilon(\vec x, t)\rangle = \check\epsilon [\beta(\vec\xi, \tau)\, , \vec u(\vec\xi ,
\tau)\, , \mu(\vec\xi , \tau)\, ;\, \vec x, t)\, .
\label{sixnine}
\end{equation}
Inverting the five relations like \eqref{sixnine} and substituting in the expressions for the
right hand side like \eqref{sixeight}, we get
\begin{equation}
\langle T^{ij}(\vec x, t)\rangle = \check T^{ij} [\langle\epsilon(\vec\xi , \tau)\rangle\, , 
\langle\vec u(\vec\xi , \tau)\rangle\, , \langle\nu(\vec\xi , \tau)\rangle\, ;\, \vec x, t)
\, .
\label{sixten}
\end{equation}
Thus, the set of equations \eqref{sixseven} can be turned into  a closed set of deterministic equations of
motion for  the expected values of the quasiclassical variables $\langle\epsilon(\vec x, t)\rangle$, 
$\langle\vec\pi(\vec x, t)\rangle$, and $\langle\nu(\vec x, t)\rangle$.

The process of expression and inversion adumbrated above could be difficult to carry out in
practice.  The familiar classical equations of motion arise from further approximations, in
particular from assuming that the gradients of all quantities are small.  For example, for a
non-relativistic fluid of particles of mass $m$,
the most general Galilean-invariant form of
the stress tensor that is linear in the gradients of the  fluid velocity  $\vec u(x)$has the approximate form \cite{LL-FM}
\begin{align}
\check T^{ij}  =&  p\delta^{ij} + m \nu \, u^i u^j
 - \eta\left[\frac{\partial u^i}{\partial x^j} + \frac{\partial u^j}{\partial x^i} - 
\frac{2}{3}\ \delta_{ij} \left(\vec\nabla \cdot\vec u\right)\right]  \nonumber \\
 &-  \zeta\, \delta_{ij} \left(\vec\nabla \cdot \vec u\right)\, .
\label{sixeleven}
\end{align}
The pressure $p$ and coefficients of viscosity $\eta$ and $\zeta$ are
themselves functions say  of $\beta$ and $\nu$ determined by the construction leading to \eqref{sixten}. This form of the stress tensor in \eqref{sixseven a} leads to the
Navier-Stokes equation.

\subsection{Quasiclassical Histories}

A quantum system can be said to behave quasiclassically when, in a suitable realm, the probability is high for its quasiclassical variables to be correlated in time by approximate, deterministic classical laws of motion. For instance, according to quantum mechanics there is a probability for the earth to move on any orbit around the sun. The earth moves classically when the probability is high that histories  of suitably coarse-grained positions of the earth's center of mass are correlated by Newton's laws of motion. The probabilities defining these correlations are probabilities of {\it time histories} of coarse-grained center of mass positions of the earth.
The behavior of the expected values of position as a function of time is not enough to evaluate these probabilities.
Similarly the  classical behavior of the expected values of quasiclassical hydrodynamic variables derived in the last subsection does not necessarily imply either the decoherence or the classicality of these variables except in very special situations. 

One example of such a special situation for the motion of  a non-relativistic particle in one dimension starts from Ehrenfest's theorem:
\begin{equation}
m \frac{d^2\langle x \rangle}{dt^2} =- \left\langle \frac{dV(x)}{dx}\right\rangle \ .
\label{C-1}
\end{equation} 
If the initial state is a narrow wave packet that does not spread very much over the time of interest this becomes approximately
\begin{equation}
m \frac{d^2\langle x \rangle}{dt^2} \approx - \frac{dV(\langle x \rangle)}{d\langle x \rangle} \ .
\label{C-2}
\end{equation} 
This is a classical equation of motion for the  expected value $\langle x \rangle$ not dissimilar in character from those for quasiclassical variables in the previous subsection. 

Suppose we study the history of the particle using a set of histories coarse-grained by ranges of $x$ at a sequence of times with the ranges all large compared to the width of the initial wave packet. Then, at sufficiently early times,  the only history with a non-negligible amplitude consists of the ranges traced out by the center of the wave packet along $\langle x(t)\rangle$. Decoherence of this set is immediate since only the one coarse-grained history tracking 
$\langle x(t)\rangle$ has any significant amplitude. And, since that history is correlated in time by \eqref{C-2}, the probablity for classical correlations in time is near unity. 

In this example the initial state is very special. Also, the coarse graining is very coarse --- too coarse for instance to exhibit any quantum corrections to classical behavior. 

In several interesting papers \cite{Hal98,Hal99, Hal03} J. Halliwell has provided a demonstration of the classical behavior of quasiclassical hydrodynamic variables. This has something of the character of the Ehrenfest example although it is more technically complex and less special and correspondingly provides  more insight into the problem of classicality. 

A very brief summary of his assumptions and results are as follows:

\begin{itemize}

\item Consider a system of $N$ non-relativistic particles in a box with a Hamiltonian $H$ specified by two-body  potentials with a characteristic range $L$. Consider an initial state $|\Psi\rangle$ that is an approximate eigenstate of the quasiclassical variables $({\epsilon}_V(\vec y,t),{\vec\pi}_V(\vec y,t), \nu_V(\vec y,t))$. (This can be achieved approximately even though these variables do not commute.) The  volume $V$ is   chosen so that  i) $\langle \nu_V(y)\rangle $ is large, and   ii) $V>>L^3$.

\item  Define a set of quasiclassical histories of the type discussed in Section IIIA, using ranges of the $({\epsilon}_V(\vec y,t),{\vec\pi}_V(\vec y,t), \nu_V(\vec y,t))$ at a sequence of times.  Under the above assumptions it is possible to show that the fluctuations in any of the quasiclassical quantities are small and remain small, for instance
\begin{equation}
\langle[ \Delta \nu_V(\vec y, t)]^2\rangle / \langle \nu_V(\vec y, t) \rangle^2 << 1 \ . 
\label{C-3}
\end{equation}
The approximate conservation underlying the quasiclassical quantities ensures that the relations like \eqref{C-3} hold over time.

\item As a consequence Halliwell shows that  histories  of the quasiclassical variables are approximately decoherent and that their probabilities are peaked about the evolution given by the classical equations of motion for the expected values. The crucial reason is that the small fluctuation relations like \eqref{C-3} mean that there is essentially only one history in each set, somewhat as in the Ehrenfest example. 

\end{itemize}

Halliwell's result is the best we have today. But in our opinion there is still much further work to be done in demonstrating the quasiclassical behavior of {\it histories} of quasiclassical variables. In particular,  as mentioned earlier,  quasiclassical realms  that are maximally refined  consistent with  decoherence and classicality are of interest. The coarse graining in the above analysis is likely to be much coarser than that. 

The contrast with the studies of classicality in the much simpler oscillator models (e.g \cite{GH93a,BH99,Hal03}) is instructive. (The reader not familiar with this work should skip this paragraph.) On the negative side these models do not deal with the quasiclassical variables under discussion here but rather  with with positions of particles or averages of these. Consequently they assume an arbitrary system/environment split. However, on the positive side it is possible to study various levels of refinement of the coarse graining, to provide quantitative estimates for such quantities as the decoherence time, and to exhibit the effects of quantum noise on classical behavior. We express the hope that it will someday be possible to achieve similar levels of precision with the more realistic quasiclassical realms.

\section{Information and Entropy}

Information must be sacrificed through coarse graining to allow the existence of probabilities for non-trivial sets of histories.  Further coarse graining is needed to achieve sets of histories that exhibit predictable regularities such as those of the quasiclassical realms. These are two messages from the previous sections. Quantitative measures of the missing information are supplied by various candidates for entropy that can be constructed in connection with realms\footnote{For discussion and comparison of some of these measures see \cite{GH90a,BH99a}.}. This section focuses on one measure of missing information and its connection with the usual entropy of chemistry and physics.

We review the general prescription for constructing entropies from coarse grainings. 
When information is missing about the state of a quantum system, that system can be described by a density matrix $\rho$. The natural and usual measure of the missing information is the
entropy of $\rho$ defined as
\begin{equation}
S(\rho) = - Tr(\rho\log \rho)\, .
\label{sevenone}
\end{equation}
This is zero when $\rho$ is a pure state, $\rho=|\Psi\rangle\langle\Psi|$, showing that a
pure state is a complete description of a quantum system.  For a Hilbert space of finite
dimension ${\cal N}$, complete ignorance is expressed by $\rho=I/Tr(I)$. The maximum value
of the missing information is $\log {\cal N}$. 

A coarse-grained description of a quantum system generally consists of specifying the
expected values of certain operators $A_m, \, (m=1, \cdots, M)$ in the ``mixed'' state $\rho$.  That is, the
system is described by specifying
\begin{equation} \langle A_m\rangle \equiv Tr(A_m\rho)\, ,\ (m=1\, , \cdots, M)\, .
\label{seventwo}
\end{equation}

For example, the $\{A_m\}$ might be an exhaustive set of orthogonal projection operators
$\{P_\alpha\}$ of the kind used in Section II to describe yes/no alternatives at one
moment of time.  These projections might be onto ranges of the  center-of-mass position of the
planet Mars. In a finer-grained quasiclassical description, of the kind discussed in
Section V, they might be projections onto ranges of values of  the densities of energy, momentum, and other conserved or nearly conserved quantities averaged over small volumes in the interior of Mars. In a 
coarse graining of that, the $\{A_m\}$ might be the operators $\epsilon_V(\vec y,t)$,
$\vec \pi_V(\vec y, t)$, and $\nu_V(\vec y, t)$ themselves, so that the description at one
time is in terms of the expected values of variables rather than expected values of projections onto ranges of values of those variables.   As these examples illustrate, the
$\{A_m\}$ are not necessarily mutually commuting. 

The measure of missing information in descriptions of the form \eqref{seventwo} for a
quantum system with density matrix $\rho$ is the maximum entropy over the effective density matrices $\tilde\rho$ that are consistent with the coarse-grained description ({\it
e.g.}~\cite{Ros83, Kat67, Jay93}). Specifically, 
\begin{equation}
S(\{A_m\},\rho) \equiv  -Tr(\tilde\rho \log\tilde\rho) \, ,
\label{seventhree}
\end{equation}
where $\tilde\rho$ maximizes this quantity subject to the constraints
\begin{equation}
Tr(A_m\tilde\rho) = \langle A_m\rangle \equiv
Tr(A_m\rho),\ m=1, \cdots,M\, .
\label{sevenfour}
\end{equation}
The solution to this maximum problem is straightforward to obtain by the method of
Lagrange multipliers and is
\begin{equation}
\tilde\rho = Z^{-1} \exp\left(-\sum^M_{m=1} \lambda^m A_m\right)\, .
\label{sevenfive}
\end{equation}
Here, the $\{\lambda_m\}$ are $c$-number Lagrange multipliers determined in terms of $\rho$ and
$A_m$ by the coarse graining constraints \eqref{sevenfour}; $Z$ ensures normalization. The
missing information is then the entropy
\begin{equation}
S(\{A_m\}, \rho) = - Tr\left(\tilde\rho \log \tilde\rho\right)\, .
\label{sevensix}
\end{equation}
As mentioned earlier the $A$'s need not be commuting for this construction.  What is
important  is that the constraints \eqref{sevenfour} are linear in $\rho$.

Suppose that  a coarser-grained description is characterized by operators $\{\bar A_{\overline m}\},
\overline{m} =1, \cdots, \overline{M}<M$ such that
\begin{equation}
\bar A_{\overline m} = \sum_{m\epsilon\overline{m}} c_{{\bar m}m} A_m\, ,
\label{sevenseven a}
\end{equation}
where the $c_{{\bar m}m}$ are  coefficients relating the operators of the finer graining to those of  the coarser graining. For example, suppose that the $A_m$ are averages over volumes $V(\vec y)$ of one of the quasiclassical variables. Averaging over bigger volumes ${\bar V}({\vec z})$ that are unions of these would be a coarser-grained description. The coefficients  $c_{{\bar m}m}$ would become  $c({{\vec z}},\vec y)\equiv V(\vec y)/{{\bar V}({\vec z})}$. 
Then, we obtain
\begin{equation}
S(\{\bar A_{\overline m}\},\rho) \geq S(\{A_m\},\rho) \, ,
\label{sevenseven b}
\end{equation}
since there are fewer constraints to apply in the maximum-missing-information construction.  Entropy
increases with further coarse graining. 

Among the various possible operators $\{A_m\}$ that could be constructed from those defining histories of the quasiclassical realm, which should we choose to find a connection
(if any), with the usual entropy of chemistry and physics?  Familiar thermodynamics provides some clues. The first law of thermodynamics connects changes in the {\it expected values} of energy over time with changes in volume and changes in entropy in accord with conservation of energy. This suggests  that for our present purpose we should consider an entropy defined at one time rather than for a history. It also suggests that we should consider the entropy defined by the expected values of quasiclassical quantities and not use  the finer-grained description by  ranges of values of the the quantities themselves, which enter naturally into histories. 

To see that this is on the right track, let us compute the missing information specifying the expected values of the total energy $H$, total momentum $\vec P$, and total conserved number $N$  for our model system in a box. The density matrix maximizing the missing information is\footnote{Note that in \eqref{seveneight} $H$ means the Hamiltonian, not the enthalpy,  and $\langle H \rangle$  is the same as what is often denoted by $U$ in thermodynamics.}, from \eqref{sevenfive}
\begin{equation}
\tilde\rho_{\rm max} = Z^{-1} \exp[-\beta\, (H-\vec U \cdot \vec P -\mu N)]\, .
\label{seveneight}
\end{equation}
This is the equilibrium density matrix \eqref{seventhree}. The entropy $S$ defined by
\eqref{sevensix} is straightforwardly calculated from the (Helmholtz) free energy $F$ defined by
\begin{align}
Tr\{\exp[-\beta\, &(H-\vec U\cdot\vec P-\mu N)]\} \nonumber \\ & \equiv \exp[-\beta(F-\vec U\cdot
\langle\vec P\rangle -\mu\langle N\rangle)]\, .
\label{sevennine}
\end{align}
We find
\begin{equation}
S=\beta(\langle H\rangle -F)\, .
\label{seventen}
\end{equation}
This standard thermodynamic relation shows that we have recovered the standard entropy of chemistry and physics. 

In an analogous way, the missing information can be calculated for a coarse graining in
which $\langle\epsilon (\vec y, t)\rangle$, $\langle\vec\pi(\vec y, t)\rangle$, and
$\langle\nu(\vec y, t)\rangle$ are specified at one moment of time.  The density matrix
that maximizes the missing information according to \eqref{seventhree} is that for  the assumed  {\it
local} equilibrium \eqref{sixfive}. The entropy is
\begin{equation}
S=\sum_{\vec y} \beta\, (\vec y, t)\, [\langle\epsilon(\vec y, t)\rangle - \langle\phi(\vec y,
t)\rangle] \, , 
\label{seveneleven}
\end{equation}
where $\langle\phi(\vec y, t)\rangle$ is the free energy density defined analogously to
\eqref{sevennine}.  The integrand of \eqref{seveneleven} is then naturally understood as  defining  the average over small volumes of an 
{\it entropy density} $\sigma(\vec x, t)$..

 What is being emphasized here is that,
in these ways, the usual entropy of chemistry and physics arises naturally from the coarse graining that is inescapable for quantum-mechanical decoherence and for quasiclassical predictability of histories, together with  the assumption of local equilibrium\footnote{In some circumstances (for example when we take gravitation into account in metastable configurations of matter) there may be other kinds of entropy  that are appropriate such as $q$-entropy with  $q \ne 1$.  Such circumstances have been excluded from our model for simplicity, but for more on them see  e.g \cite{GT04}}.

\section{The Second Law of Thermodynamics}

We can now discuss the time evolution in our universe of the entropy described in the previous section by the coarse graining defining a quasiclassical realm.  Many pieces of this  discussion are standard (see, e.g. \cite{Zub74})  although not often described in the context of quantum cosmology.  

The context of this discussion continues to be our model universe of fields in a box. For the sake of simplicity, we are not only treating an artificial universe in a box but also sidestepping vital issues such as the accelerated expansion of the universe, possible eternal inflation, the decay of the proton, the evaporation of black holes,  long range forces, gravitational clumping, etc. etc. (See, for example \cite{Alb03,Vil83,Lin86}.) 

For this discussion we will distinguish two connected but different features of the universe:
\begin{itemize}
\item The tendency of the total entropy\footnote{ In the context of our model universe in a box the total entropy is defined. For the more realistic cases of a flat or open universe the total entropy may be infinite and  we should refer to the entropy of a large comoving volume.} of the universe to increase. 
\item The tendency of the  entropy of each  presently almost isolated system to increase in the same direction of time. This might be called the homogeneity of the thermodynamic arrow of time. 
\end{itemize}  
Evidently these features are connected. The first follows from the second, but only in the late universe when almost isolated systems are actually present.  In the early universe we have only the first. Together they may be called the second law of thermodynamics. 

More particularly,  isolated systems described by quasiclassical coarse grainings based on approximately
conserved quantities can be expected to evolve toward equilibrium characterized by the total values of these
conserved quantities. In our model universe in a box, the probabilities of
$\epsilon_V(\vec y, t)$, $\vec \pi_V(\vec y, t)$, $\nu_V(\vec y, t)$, and their
correlations are eventually given by the equilibrium density matrix \eqref{sixthree}. The
conditions that determine the equilibrium density matrix are sums of the conditions that
determine  the local equilibrium density matrix in the Jaynes construction \eqref{sevenfour}. The smaller number of conditions means that the equilibrium entropy will be larger than that for any local equilibrium [{\it cf.}~\eqref{sevenseven b}]. 

Two conditions are necessary for our universe to exhibit a general increase in total entropy defined by quasiclassical variables:
\begin{itemize}
\item The quantum state $|\Psi\rangle$ is such that the initial entropy is near the minimum it could have for the coarse graining defining it. It then has essentially nowhere to go but up. 
\item The relaxation time to equilibrium is long compared to the present age of the universe so that the general tendency of its entropy to increase will dominate its evolution.
\end{itemize}  

In our simple model we have neglected gravitation for simplicity, but for the following discussion we restore it. 
Gravity is essential to realizing the first of these conditions because in a self-gravitating system gravitational clumping increases entropy. The early universe is approximately homogeneous, implying that the entropy has much more room to increase through the gravitational growth of fluctuations. In a loose sense, as far as gravity is concerned, the entropy of the early universe is low for the coarse graining we have been discussing.  The entropy then increases. Note that a smaller effect in the opposite direction is connected with the thermodynamic equilibrium of the matter and radiation in the very early universe. See \cite{LB90} for an entropy audit of the {\it present} universe.

Coarse graining by approximately conserved quasiclassical variables helps with the second of the two conditions above --- relaxation time short compared to present age. 
Small volumes come to local equilibrium quickly. But the approximate conservation ensures that the whole system will approach equilibrium slowly, whether or not such equilibrium is actually attained.  Again gravity is important because its effects concentrate a significant fraction of the matter into almost  isolated systems such as stars and galactic halos, which strongly  interact with one other only infrequently\footnote{For thoughts on what happens in the very long term in an expanding universe see \cite{longterm,AL97}.}.

Early in the universe there were no almost isolated systems. They arose later  from the condensation of initial fluctuations by the action of gravitational attraction\footnote{Much later certain IGUSes create isolated systems in the laboratory which typically inherit the thermodynamic arrow of the IGUS and apparatus that prepared them.}. They are mostly evolving toward putative equilibrium in the same direction of time. 
Evidently this homogeneity of the thermodynamic arrow of time cannot follow from the approximately time-reversible dynamics
and statistics alone. Rather the explanation is that the progenitors of today's nearly isolated
systems were all far from equilibrium  a long time ago and have been running down hill ever
since.  This provides a stronger constraint on the initial state than merely having low total entropy.  As Boltzmann put it
over a century ago: ``The second law of thermodynamics can be proved from the
[time-reversible] mechanical theory, if one assumes that the present state of the
universe\dots started to evolve from an improbable [{\it i.e.}~special] state''
\cite{Bol97}. 

The initial quantum state of our universe must be such that it leads to the decoherence of
sets of quasiclassical histories  that describe coarse-grained spacetime geometry and matter fields. Our observations require this now, and the successes of the classical history of the universe suggests that there was  a quasiclassical realm at a very early time. In addition,  the initial state must 
be such that the entropy of quasiclassical coarse graining is low in the beginning and also be such  that the entropy of presently isolated systems was also low. Then the universe can exhibit both aspects of the second law of thermodynamics.

The quasiclassical coarse grainings are therefore distinguished from others, not only
because they exhibit predictable regularities of the universe governed by approximate
deterministic equations of motion, but also because they are characterized by a sufficiently  low
entropy in the beginning and a slow evolution towards equilibrium,  which makes those regularities exploitable.

This confluence of features suggests the possibility of a  connection between the dynamics of the
universe and its initial condition. The `no-boundary' proposal for the initial quantum
state \cite{HH83} is an example of just such a connection. According to it, the initial state is computable from the Euclidean action, in a way similar to the way the ground state of a system in flat space can be calculated.

\section{Conclusions and Comments}

Many of the conclusions of this paper can be found among the bullets in the Introduction. Rather than reiterating all of them here, we prefer to discuss in this section some broader issues. These concern the  relation between our approach to quantum mechanics, based on coarse-grained decoherent histories of a closed system, and the approximate quantum mechanics of measured subsystems, as in the ``Copenhagen interpretation.''
The latter formulation {\it postulates} (implicitly for most authors or explicitly in the case of Landau and Lifshitz \cite{LL58}) a classical world and a quantum world, with a movable boundary  between the two. Observers and their measuring apparatus make use of the classical world, so that  the results of  a ``measurement'' are  ultimately expressed in one or more ``c-numbers''. 

We have emphasized that this widely taught interpretation, although successful, cannot be the fundamental one because it seems to require a physicist outside the system making measurements (often repeated ones) of it. That would seem to rule out any application to the universe, so that quantum cosmology would be excluded. Also billions of years went by with no physicist in the offing. Are we to believe that quantum mechanics did not apply to those times?

In this discussion, we will concentrate on how the Copenhagen approach fits in with ours as a set of special cases and how the ``classical world'' can be replaced by  a quasiclassical realm. Such a realm is not {\it postulated} but rather is {\it explained} as an emergent feature of the universe characterized by $H$, $|\Psi\rangle$, and the enormously long sequences of accidents (outcomes of chance events) that constitute the coarse-grained decoherent histories. The material in the preceding sections can be regarded as a discussion of how  quasiclassical realms emerge.

We say that a `measurement situation' exists if some variables (including such quantum-mechanical variables as electron spin) come into high correlation with a quasiclassical realm.  In this connection we have often referred to fission tracks in mica. Fissionable impurities can undergo radioactive decay and produce fission tracks with randomly distributed  definite directions. The tracks are there irrespective of the presence of an ``observer''. It makes no difference if a physicist or other human or a chinchilla or a cockroach looks at the tracks. Decoherence of the alternative tracks induced by interaction with the other variables in the universe is what allows the tracks to exist independent of ``observation'' by an ``observer''. All those other variables are effectively doing the ``observing''. The same is true of the successive positions of the moon in its orbit  not depending on the presence of ``observers'' and for density fluctuations in the early universe existing when there were no observers around to measure them.

The idea of ``collapse of the wave function'' corresponds to the notion of variables coming into high correlation with a quasiclassical realm, with its decoherent histories that give true probabilities. The relevant histories are defined only through the projections that occur in the expressions for these probabilities [cf \eqref{twosix}]. Without projections, there are no questions and no probabilities.   In many cases conditional probabilities are of interest.  The collapse of the probabilities  that occurs in their construction is no different from the collapse that occurs at a horse race when a particular horse wins and future probabilities for further races conditioned on that event become relevant.

The so-called ``second law of evolution'',  in which a state is `reduced' by the action of a projection, and the probabilities renormalized to give ones conditioned on that projection, is thus not some mysterious feature of the measurement process.  Rather it is a natural consequence of the quantum mechanics of decoherent histories, dealing with alternatives  much more general than mere measurement outcomes. 

There is thus no actual conflict between the Copenhagen formulation of quantum theory and the more general quantum mechanics of closed systems. Copenhagen quantum theory is an approximation to the more general theory that is appropriate for the special case of measurement situations. Decoherent histories quantum mechanics is rather a {\it generalization} of the usual approximate quantum mechanics of measured subsystems.

In our opinion decoherent histories quantum theory advances our understanding in the following ways among many others: 

\begin{itemize}
\item Decoherent histories quantum mechanics extends the domain of applicability of quantum theory to histories of features of the universe irrespective of whether they are receiving attention of observers and in particular to histories describing the evolution of the universe in cosmology.  
 
\item The place of classical physics in a quantum universe is correctly understood as a property of a particular class of sets of  decoherent coarse-grained alternative histories --- the quasiclassical realms \cite{GH93a,Har94b}.
In particular, the {\it limits} of a quasiclassical description can be explored. Dechoherence may fail if the graining is too fine. Predictability is limited by quantum noise and by the major branchings that arise from the amplification of quantum phenomena as in a measurement situation. Finally, we cannot expect a quasiclassical description of the universe in its earliest moments where the very geometry of spacetime may be undergoing large quantum fluctuations. 

\item Decoherent histories quantum mechanics provides new connections such as the relation (which has been the subject of this paper)  between the coarse graining characterizing quasiclassical realms and the coarse graining characterizing the  usual thermodynamic entropy of chemistry and physics. 

\item Decoherent histories quantum theory helps with understanding  the Copenhagen approximation. For example, measurement was characterized as an ``irreversible act of amplification'', ``the creation of a record'', or as ``a connection with macroscopic variables''. But these were inevitably imprecise ideas. How much did the entropy have to increase, how long did the record have to last, what exactly was meant by ``macroscopic''? Making these ideas precise was a central problem for a theory in  which measurement is fundamental. But it is less central in a theory where measurements are just special, approximate situations among many others. 
Then characterizations such as those above are not false, but true in an approximation that need not be exactly defined.

\item Irreversibility clearly plays an important role in science as illustrated here by the two famous applications to quantum-mechanical measurement situations and to thermodynamics. It is not an absolute concept but context-dependent like so much else in quantum mechanics and statistical mechanics. It is highly  dependent on coarse graining, as in the case of the document shredding \cite{Gel94}. This was typically carried out in one dimension until the seizure by Iranian ``students'' of the U.S. Embassy in Tehran in 1979, when classified documents were put together and published. Very soon, in many parts of the world, there was a switch to two-dimensional shredding, which still appears to be secure today.
 It would now be labeled as irreversible just as the  one-dimensional one was previously. The shredding and mixing of shreds clearly increased the entropy of the documents, in both cases by an amount dependent on the coarse grainings involved. Irreversibility is evidently not absolute but dependent on the effort or cost involved in reversal.
 
\end{itemize}. 

The founders of quantum mechanics were right
in pointing out that something external to the framework of wave function
and Schr\"odinger equation {\it is} needed to interpret the theory.  But
it is not a postulated classical world to which quantum mechanics does not
apply.  Rather it is the initial condition of the universe that, together
with the action function of the elementary particles and the throws of quantum
dice since the beginning, explains the origin of quasiclassical realm(s)
within quantum theory itself.

\appendix

\renewcommand{\theequation}{\Alph{section}.\arabic{equation}}

\section{Trivial Decoherence of Perfectly Fine-Grained Sets of Histories}

A set of perfectly fine-grained histories trivially decoheres when there is a different  final projection for each history with a non-zero branch state vector.  Equivalently,
we could say that a set of completely fine-grained histories is trivial if each final
projection has only one possible prior history.  The orthogonality of the
final alternatives then automatically guarantees the decoherence of the realm. Such sets are trivial in
the sense that what happens at the last moment is uniquely correlated with the
alternatives at all previous moments. This appendix completes the demonstration, adumbrated in Section III.A, that completely fine-grained realms are trivial.

The domain of discussion was described in Section III.A.  We consider sets of histories composed of completely fine-grained alternatives at any one time specified by a complete basis for Hilbert space. It is sufficient to consider a sequence of times $t_k, (k=1, 2, \cdots, n)$  because if such decoherent sets are trivial all finer-grained sets will also be trivial. In Section III.A we denoted the bases by $\{|k, i_k\rangle\}$. Here, in the hope of keeping the
notation manageable, we will drop the first label and just write $\{|i_k\rangle\}$. The
condition for decoherence \eqref{twoseven} is then

\begin{equation}
\left\langle \Psi_{i^\prime_n,\cdots, i^\prime_1}|\Psi_{i_n,\cdots, i_1}\right\rangle =
p(i_n,\cdots, i_1)\, \delta_{i^\prime_ni_n}\cdots \delta_{i^\prime_1i_1}
\label{aone}
\end{equation}
where the branch state vectors $|\Psi_{i_n,\cdots, i_1}\rangle$ are defined by
\eqref{twofive} with alternatives at each time of the form \eqref{threeone} and where equality has replaced $\approx$ because we are insisting on exact decoherence. 

Mathematically, the decoherence condition \eqref{aone} can be satisfied in several
ways. Some of the histories could be represented by branch state vectors which vanish
(zero histories for short). These have vanishing inner products with all branch state vectors including themselves. They therefore do not affect decoherence and their probabilities are equal to zero.  Further discussion can therefore be restricted to the
non-vanishing branches.

Decoherence in the final alternatives $i_n$ is automatic because the projections
$P_{i^\prime_n}$ and $P_{i_n}$ are orthogonal if different. Decoherence is also automatic if each non-zero history has its own final alternative or, equivalently, if each final
alternative $i_n$ has a unique prior history. Then decoherence of the final alternatives 
guarantees the decoherence of the set.

Two simple and trivial examples may help to make this discussion concrete.  Consider
the set of completely fine-grained histories defined by taking the same basis
$\{|i\rangle\}$ at each time. 
The
resulting orthogonality of the projections between times ensures that the only non-zero
histories have $i_n=i_{n-1}=\cdots=i_2=i_1$ in an appropriate notation.   Evidently each $i_n$ corresponds to
exactly one chain of past alternatives and the decoherence condition \eqref{aone} is
satisfied.

A related trivial example is obtained choosing the $\{|i_k\rangle\}$ to be the state
$|\Psi\rangle$ and some set of orthogonal states at each of the times $t_1, \cdots,
t_{n-1}$.  That is, we are  mindlessly asking the question, ``Is the system in the state
$|\Psi\rangle$ or not?'' over and over again until the last alternative. The only
non-zero branches are of the form
\begin{equation}
|i_n\rangle\langle i_n|\Psi\rangle\langle\Psi|\Psi\rangle \cdots
\langle\Psi|\Psi\rangle=|i_n\rangle\langle i_n|\Psi\rangle\, .
\label{atwo}
\end{equation}
In this case, each final alternative is correlated with the same set of previous
alternatives.  The set decoheres because the only non-trivial branching is at the last
time as the equality in \eqref{atwo} shows.  But then each history has a unique final
end alternative and the set  is thus  trivial\footnote{Coarse-grained examples of both kinds of triviality can be given. Repeating the same Heisenberg picture sets of projections at all times is an example of the first. Histories defined by Heisenberg picture  sets of  projections onto a subspace containing $|\Psi\rangle$ and other orthogonal subspaces are examples of the second.}

Trivial sets of completely fine-grained histories have many zero-histories as in the above
examples. To see this more quantitatively, imagine for a moment that the dimension of
Hilbert space is a very large but finite number $\cal N$.  A generic, completely
fine-grained set with $n$ times would consist of ${\cal N}^n$ histories. But there can be at
most $\cal N$ orthogonal branches --- no more than are be supplied just by alternatives at
one time. Indeed the trivially decohering completely fine-grained sets described above
have at most $\cal N$ branches. Most of the possible histories must therefore be zero.
Further, assuming that at least one non-zero branch is added at each time, the number of
times $n$ is limited to $\cal N$. 

Were a final alternative in a completely fine-grained set to have more than one
possible previous non-zero history, the set would not decohere. To see this, suppose
some particular final alternative $k_n$ had two possible, non-zero, past histories. Let the
earliest (and possibly the only) moment the histories differ be time $t_j$. The
condition for decoherence is [{\it cf.} \eqref{aone}]
\begin{align}
\langle\Psi|i^\prime_1\rangle\langle i^\prime_1|i^\prime_2\rangle&\cdots\langle
i^\prime_{n-1}|k_n\rangle\langle k_n|i_{n-1}\rangle \cdots  \langle i_2|i_1\rangle
\langle i_1|\Psi\rangle \nonumber \\&= p(k_n, \cdots, i_1)\, \delta_{i^\prime_{n-1} i_{n-1}}\cdots
\delta_{i^\prime_1 i_1}\, .
\label{athree}
\end{align}
Sum this over all the $i_1, \cdots, i_{j-1}$ and  $i'_1, \cdots, i'_{j-1}$  to find
\begin{align}
\langle\Psi|i^\prime_j\rangle\, &M\left(i^\prime_j, \cdots, i^\prime_{n-1}, k_n, i_{n-1},
\cdots, i_j\right)\, \langle i_j|\Psi\rangle \nonumber \\  &= p(k_n, \cdots, i_j)
\, \delta_{i^\prime_{n-1} i_{n-1}}\cdots \delta_{i^\prime_j i_j}
\label{afour}
\end{align}
where we have employed an abbreviated notation for the product of all the remaining
matrix elements.  By hypothesis there are at least two chains $(i_j, \cdots, k_n)$
differing in the value of $i_j$ for which $M$ does not vanish and the history is not
zero. But decoherence at time $t_j$ then requires that $\langle i_j|\Psi\rangle$ vanish
for all but one $i_j$,  contradicting the assumption that there are two non-zero
histories. 

The only decohering, perfectly  fine-grained sets are thus trivial.

We now discuss the connection between this notion of a trivial fine-grained realm and the idea of  generalized records introduced in \cite{GH90b,
GH95}.  Generalized records of a realm are a set of orthogonal  projections $\{R_\alpha\}$ that are correlated to a good approximation with the histories of the realm $\{ C_\alpha \}$ so that
\begin{equation}
R_\alpha C_\beta|\Psi\rangle \approx \delta_{\alpha\beta} C_\alpha |\Psi\rangle \, .
\label{afive}
\end{equation} 
Generalized records of a particular kind characterize important physical mechanisms of decoherence. For instance, in the classic example of Joos and Zeh \cite{JZ85} a dust grain is imagined to be initially in a superposition of two places deep in intergalactic space. Alternative histories of coarse-grained positions of the grain are made to decohere by  the interactions with the $10^{11}$ photons of the $3^\circ$  cosmic background radiation that scatter every second. Through the interaction, records of the positions  are created in the photon degrees of freedom, which are variables {\it not} followed by the coarse graining.  

But the trivial generalized records that characterize fine-grained realms are not of this kind. They  are in variables that {\it are}  followed by the coarse graining. They could not be in other variables because the histories are fine-grained and follow everything.  They are trivial records for a trivial kind of realm\footnote{In the general case of medium decoherence, we can construct projection operators onto orthogonal spaces each of which includes one of the states $C_\alpha|\Psi\rangle$ \cite{GH95}, but again these operators are not what we are really after in the way of generalized records {\it if they deal mainly with what is followed.}}.

\section{No Non-trivial  Certainty Arising from the Dynamics Plus the Initial Condition}

The Schr\"odinger equation is deterministic and the evolution of the state vector is
certain.  In this appendix we show that this kind of determinism is the only source of  histories that  are certain (probability equal $1$), 
coarse-grained or not, for a closed quantum-mechanical system. 

Suppose we have a set of exactly decoherent histories of the form \eqref{twothree}, one
member of which has probability 1, {\it i.e.}~the history is certain.  Denote the class 
operator for this particular history by $C_1$ and index its particular alternatives at 
each time so they are
$\alpha_1=\alpha_2=\cdots \alpha_n=1$. Thus we have
\begin{equation}
p(1)=\|C_1|\Psi\rangle\|^2=1\ ,\ p(\alpha)=\|C_\alpha |\Psi\rangle\|^2=0
\ , \ \alpha\not= 1\, .
\label{threesix}
\end{equation}
The second of these implies $C_\alpha|\Psi\rangle=0$ for $\alpha\not=1$. Given that
$\Sigma_\alpha C_\alpha=I$, the first implies that $C_1|\Psi\rangle=|\Psi\rangle$.
In summary we can write
\begin{align}
C_\alpha |\Psi\rangle =& P^1_{\alpha_n} (t_n) \cdots P^1_{\alpha_1}(t_1)|\Psi\rangle \nonumber \\ =
&\delta_{\alpha_n, 1}\delta_{\alpha_{n-1}, 1}\cdots \delta_{\alpha_1, 1} |\Psi\rangle
\, .
\label{threeseven}
\end{align}
Summing both sides over this over all $\alpha$'s except those at time $t_k$, we get
\begin{equation}
P^k_{\alpha_k}(t_k)|\Psi\rangle = \delta_{\alpha_k, 1}|\Psi\rangle\, .
\label{threeeight}
\end{equation}
This means that all projections in a set of histories of which one is certain are either
onto subspaces which contain $|\Psi\rangle$ or onto subspaces orthogonal to $|\Psi\rangle$ .  Effectively they
correspond to questions that ask, ``Is the state still $|\Psi\rangle$ or not?''. That  
 is how the certainty of unitary evolution is represented in the Heisenberg picture,
where $|\Psi\rangle$ is independent of time.

Note that this argument doesn't exclude histories which are essentially
certain over big stretches of
time as in histories describing alternative values of a conserved quantity at many times.
Then there would be, in general, initial probabilities for the value of the conserved
quantity which then do not change through subsequent history.

\acknowledgments
We thank J. Halliwell and S. Lloyd for useful recent discussions. 
We thank the Aspen Center for Physics for hospitality over several summers while this work was in progress. JBH thanks the Santa Fe Institute for supporting several visits there. The work of JBH was supported in part by the National Science Foundation under grant PHY02-44764 and    PHY05-55669.
 The work of MG-M was supported by the C.O.U.Q. Foundation,  by Insight Venture Management, and by the KITP in Santa Barbara.  The generous help provided by these organizations is gratefully acknowledged. 
 


\begin{thebibliography}{99}

\bibitem{GL04}  M.~Gell-Mann and S.~Lloyd, {\it Effective Complexity}  in {\sl Non-Extensive Entropy ---Interdisciplinary Applications}, ed. by. M.~Gell-Mann and C. Tsallis,
Oxford University Press, New York (2004). 

\bibitem{GH90a} M.~Gell-Mann and J.B.~Hartle, {\it Quantum Mechanics in the
Light of Quantum Cosmology}, in {\sl Complexity, Entropy, and the Physics
of Information, SFI Studies in the Sciences of Complexity}, Vol.~VIII,
ed.~by W.~Zurek,  Addison Wesley, Reading, MA (1990).

\bibitem{GH90b} M.~Gell-Mann and J.B.~Hartle, {\it Alternative Decohering
Histories in Quantum Mechanics}, in the {\sl Proceedings of
the 25th International Conference on High Energy Physics}, Singapore,
August, 2-8, 1990, ed.~by K.K.~Phua and Y.~Yamaguchi (South East Asia
Theoretical Physics Association and Physical Society of Japan) distributed
by World Scientific, Singapore (1990).

\bibitem{Har91a} J.B.~Hartle, {\it The Quantum Mechanics of Cosmology}, in {\sl
Quantum Cosmology and Baby Universes:  Proceedings of the 1989 Jerusalem Winter
School for Theoretical Physics}, ed.~by ~S.~Coleman, J.B.~Hartle, T.~Piran,
and S.~Weinberg, World Scientific, Singapore (1991), pp. 65-157.

\bibitem{GH93a} M.~Gell-Mann and J.B.~Hartle, {\it Classical Equations for
Quantum Systems}, {\sl Phys.~Rev.~D}, {\bf 47}, 3345 (1993);
gr-qc/9210010.

\bibitem{GH93b}  M.~Gell-Mann and J.B.~Hartle, {\it Time Symmetry and
Asymmetry in Quantum Mechanics and Quantum Cosmology}, in {\sl The Physical
Origins of Time Asymmetry}, ed.~by J.~Halliwell, J.~P\'erez-Mercader, and W.~Zurek,
Cambridge University Press, Cambridge (1994); gr-qc/9309012.

\bibitem{Gel94} M.~Gell-Mann, {\sl The Quark and the Jaguar}, W.H..~Freeman
New York (1994).

\bibitem{GH94} M.~Gell-Mann and J.B.~Hartle, {\it Equivalent Sets of
Histories and Multiple Quasiclassical Domains}; gr-qc/9404013.

\bibitem{Har91b} J.B.~Hartle, {\it Spacetime Coarse Grainings in
Non-Relativistic Quantum
Mechanics}, {\sl Phys. Rev., D}  {\bf 44}, 3173 (1991).

\bibitem{Ishsum} C.J.~Isham, {\it Quantum Logic and the Histories Approach to Quantum Theory}, {\sl J.~Math.~Phys.}, {\bf 35}, 2157
(1994); C.J.~Isham and N.~Linden,{\it Quantum Temporal Logic in the Histories Approach to Generalized Quantum Theory}.  {\sl J.~Math.~Phys.}, {\bf 35}, 5452(1994). 

\bibitem{Har94b} J.B.~Hartle, {\it Quasiclassical Domains In A Quantum Universe},
 in {\sl Proceedings of the Cornelius  Lanczos International
 Centenary Conference}, ed.~by J.D.~Brown, M.T.~Chu, D.C.~Ellison, R.J.~Plemmons,
SIAM, Philadelphia, (1994); gr-qc/9404017.

\bibitem{GH95}  M.~Gell-Mann and J.B.~Hartle, {\it Strong Decoherence},
in the {\sl Proceedings of the 4th Drexel Symposium on
Quantum Non-Integrability --- The Quantum-Classical Correspondence},
Drexel University, September 8-11, 1994, ed.~by. D.-H.~Feng and B.-L.~Hu,
International Press, Boston/Hong-Kong (1995); gr-qc/9509054.

\bibitem{Gri02} R.B.~Griffiths, {\sl Consistent Quantum Theory}, Cambridge
University Press, Cambridge, UK (2002).

\bibitem{Omn94} R.~Omn\`es, {\sl Interpretation of Quantum Mechanics},
Princeton University Press, Princeton (1994).
 
\bibitem{Har95c} J.B.~Hartle, {\it Spacetime Quantum Mechanics and the
Quantum Mechanics of Spacetime} in {\sl Gravitation and Quantizations},
in the {\sl Proceedings of the 1992 Les Houches Summer School},
ed.~by B.~Julia and J.~Zinn-Justin, Les Houches Summer
School Proceedings, Vol.~LVII, (North Holland, Amsterdam, 1995);
gr-qc/9304006. A pr\'ecis of these lectures is given in {\it Quantum Mechanics at the Planck
Scale}, talk given at the {\sl Workshop on Physics at the Planck Scale}, Puri,
India, December 1994;  gr-qc/9508023.

\bibitem{Har06}  J.B.~Hartle, {\sl Generalizing Quantum Mechanics for Quantum Spacetime}, to appear in the proceedings of the 23rd Solvay Conference, {\sl The Quantum Structure of Space and Time}; gr-qc/0602013.

\bibitem{Dio04}  L.~Di\'osi, {\it Anomalies of Weakened Decoherence Criteria for Quantum Histories}, {\sl Phys.~Rev.~Lett.}, {\bf 92}, 170401 (2004). 

\bibitem{Har04} J.B.~Hartle, {\it Linear Positivity and Virtual Probability},
{\sl Phys.~Rev.~A}, {\bf 70}, 022104 (2004); quant-ph/0401108.


\bibitem{JZ85}  E.~Joos and H.D.~Zeh, {\it The Emergence of Classical Properties Through
Interaction with the Environment}, {\sl Zeit. Phys.~B}, {\bf 59}, 223 (1985).

\bibitem{DK96} H.F.~Dowker and A.~Kent, {\it On the Consistent Histories
Approach to Quantum Mechanics}, {\sl J.~Stat.~Phys.} {\bf 82}, 1574, (1996);
gr-qc/9412067.

\bibitem{Dio95} L.~Diosi, {\it On the Maximum Number of Decoherent Histories}, {\sl Phys. Lett.}, {\bf 203}, 267-268 (1995). 

\bibitem{Cra97}  D.~Craig, {\it The Geometry of Consistency:  Decohering Histories in Generalized Quantum Theory}, gr-qc/9704031. 

\bibitem{Hal98} J.~Halliwell, {\it Decoherent Histories and Hydrodynamic Equations},
{\sl Phys.~Rev.~D}, {\bf 58}, 1050151--12 (1998); quant-ph/9805062.

\bibitem{Hal99}J.~Halliwell, {\it  Decoherent Histories and the Emergent Classicality of Local Densities}, {\sl  Phys. Rev. Lett.}  {\bf 83},  2481 (1999).


\bibitem{Hal03} J.~Halliwell, {\it Decoherence of Histories and Hydrodynamic Equations
for Linear Oscillator Chain}, {\sl Phys.~Rev.~D}, {\bf 68}, 025018 (2003);
quant-ph/0305084.

\bibitem{FV63}  R.P.~Feynman and J.R.~Vernon, {\it The Theory of a General Quantum System
Interacting with a Linear Dissipative System}, {\sl Ann.~Phys.~(N.Y.)}
{\bf 24}, 118--173 (1963).

\bibitem{CL83} A.~Caldeira  and A.~Leggett, {\it Path Integral Approach to Quantum
Brownian Motion}, {\sl Physica A} {\bf 121}, 587 (1983).

\bibitem{BH99} T.~Brun and J.B.~Hartle, {\it Classical Dynamics of the
Quantum Harmonic Chain}, {\sl Phys.~Rev.~D}, {\bf 60}, 123503 (1999);
quant-ph/9905079. 

\bibitem{Fos75} D.~Forster, {\sl Hydrodynamic Fluctuations, Broken Symmetry,
and Correlation Functions}, Addison-Wesley, Redwood City, (1975).

\bibitem{Zub74} D.N.~Zubarev, {\sl Nonequilibrium Statistical
Thermodynamics}, \ed P.~Gray and P.J.~Shepherd, Consultants Bureau, New
York, (1974).

\bibitem{LL-FM} L.~Landau and E.~Lifshitz, {\sl Fluid Mechanics}, Pergamon, London (1959). 

\bibitem{BH99a} T.~Brun and J.B.~Hartle, {\sl  Entropy of Classical Histories},
{\sl Phys.~Rev.~E}, {\bf 59}, 6370-6380 (1999); gr-qc/9808024. 

\bibitem{Ros83} E.T.~Jaynes, {\sl Papers on Probability Statistics and 
Statistical Mechanics}, ed.~by R.D.~Rosenkrantz, D.~Reidel, Dordrecht (1983).

\bibitem{Kat67} A.~Katz, {\sl Principles of Statistical Mechanics: The Information 
Theory Approach}, W.H.~Freeman, San Francisco, (1967).

\bibitem{Jay93} E.T.~Jaynes, {\sl Probability Theory}, Cambridge University Press,
Cambridge, UK (1993).

\bibitem{Vil83} A.~Vilenkin, {\it Quantum Fluctuations in the New Inflationary Universe},  {\sl Nucl. Phys. B}, {\bf 226}, 527 (1983).

\bibitem{Lin86} A.~Linde, {\it  Eternal Chaotic Inflation},  {\sl Mod. Phys. Lett. A}, {\bf 1} 81-85, (1986);  A. Linde, {\it Eternally Existing Self-reproducing Chaotic Inflationary Universe}. {\sl Phys. Lett. B} {\bf 175} 395-400 (1986).

\bibitem{Alb03}  A.~Albrecht, {\it Cosmic Inflation and the Arrow of Time}
in {\sl Science and Ultimate Reality: From Quantum to Cosmos, honoring John Wheeler's 90th birthday} ed. by J. D. Barrow, P.C.W. Davies, and C.L. Harper,  Cambridge University Press, Cambridge (2003); astro-ph/0210527.

\bibitem{LB90} B.~Basu and D. Lynden-Bell, {\it A Survey of Entropy in the Universe},
{\sl QJRAS}, {\bf 31}, 359-369 (1990). 

\bibitem{longterm} F. Dyson, {\it Time Without End: Physics and Biology in an Open Universe},  {\sl Rev. Mod. Phys.}, {\bf 51}, 447 (1979).

\bibitem{AL97}  F.C.~Adams and G.~Laughlin, {\it A Dying Universe: The Long-term Fate and Evolution of Astrophysical Objects}, {\sl Rev. Mod. .Phys.}, {\bf 69}, 337 (1997). 

\bibitem{Bol97} L.~Boltzmann, {\it Zu Hrn.~Zermelo's Abhandlung \"Uber
die mechanische Erkl\"arung Irreversibler Vorgange}, {\sl Ann.~Physik},
{\bf 60}, 392-398 (1897).


\bibitem{HH83} J.B.~Hartle and S.W.~Hawking, {\it Wave Function of the
Universe}, {\sl Phys.~Rev.~D}, {\bf 28}, 2960 (1983).

\bibitem{LL58} L.~Landau and E.~Lifshitz, {\sl Quantum Mechanics},
Pergamon, London (1958).

\bibitem{GT04} C.~Tsallis, {\it Possible Generalizations of Boltzmann-Gibbs Statistics}, {\sl J. Stat. Phys.}, {\bf 52}, 479 (1988);  C.~Tsallis, M. Gell-Mann, and Y.~Sato, {\it Asymptotically Scale-Invariant Occupancy of Phase Space Makes the q-Entropy Extensive}, {\sl Proc. Nat. Acad. Sci. (USA)}, {\bf 153}, 15382 (2005); S. Umarov, C. Tsallis, and S. Steinberg,
{\it  A Generalization of the Central Limit Theorem Consistent with Non-Extensive Statistical Mechanics}, {\sl Ann. Probability}, (submitted); S. Umarov, C. Tsallis, M. Gell-Mann and S. Steinberg, {\it q-Generalization of Symmetric $\alpha$-Stable Distributions}, Parts I and II, {\sl Ann. Probability}, (submitted).




\end{thebibliography}
\end{document}